\documentclass[preprint,12pt]{elsarticle}

\usepackage{amsmath,amssymb}
\usepackage{graphicx}
\usepackage{hyperref}
\usepackage{xcolor}

\journal{Physica A: Statistical Mechanics and its Applications}

\begin{document}

\begin{frontmatter}

\title{Fidelity susceptibility and geometric response in flux-tuned Dirac
systems: exact results from a low-energy two-level reduction}

\author{C. A. S. Almeida}
\address{Universidade Federal do Cear\' {a} (UFC), Departamento de F\'{i}sica,
Campus do Pici, Fortaleza-CE, 60455-760, Brazil.}
\ead{carlos@fisica.ufc.br}

\begin{abstract}
We study the parametric sensitivity of two-dimensional massive Dirac
fermions subject to Aharonov--Bohm flux insertion, using the Bures metric
(fidelity susceptibility) as the central statistical-mechanical response
function. Near integer values of the reduced flux, the low-energy spectrum
undergoes a flux-induced avoided crossing whose structure is controlled by
the Dirac mass. Through a controlled low-energy projection of the full
Dirac-Aharonov-Bohm operator onto an effective two-level subspace --
valid in the vicinity of integer flux values -- we derive an exact
closed-form expression for the ground-state Bures metric, which takes a
universal Lorentzian profile centered at integer flux values with width set
by the mass parameter. The peak value scales as $g^{\max}_{\lambda\lambda}
\sim m^{-2}$, diverging in the chiral limit in direct analogy with the
divergence of thermodynamic susceptibilities near critical points. We
introduce an integrated geometric susceptibility $\chi(m) = \pi/(8m)$,
whose inverse-mass scaling is the information-geometric counterpart of
power-law critical behavior, with the Dirac mass playing the role of a
relevant coupling controlling the distance from the chiral fixed point.
The Lorentzian profile is shown to arise from the curvature of the
ground-state manifold on the Bloch sphere, requiring no dynamical input
beyond the spectral structure. Importantly, this geometric response is
independent of Berry curvature and topological invariants, emerging instead
from a universal local spectral mechanism. Through its spectral
representation, the Bures metric is identified as the geometric
(paramagnetic) contribution to the persistent current susceptibility,
encoding the sensitivity of persistent currents and orbital magnetization
to flux variations and establishing a direct connection between information
geometry and physically measurable response functions in mesoscopic Dirac
systems.
\end{abstract}


\end{frontmatter}

\section{Introduction}

The geometry of quantum states provides a natural framework for characterizing
the parametric sensitivity of quantum systems, with deep connections to statistical
mechanics and thermodynamics. Among the available geometric tools, the Bures
metric --- equivalently, the fidelity susceptibility for pure states --- endows the
space of control parameters with a Riemannian structure and quantifies the
distinguishability between neighboring quantum states~\cite{Provost1980,
Petz1996, Bengtsson2017}. In the language of statistical mechanics, the fidelity
susceptibility plays a role analogous to thermodynamic response functions: it
measures how sensitively the ground-state structure responds to external
perturbations, diverging near points of enhanced spectral sensitivity in a manner
reminiscent of susceptibility scaling near critical points~\cite{Zanardi2006,
Gu2010}.

Pronounced geometric responses frequently arise in the vicinity of avoided level
crossings. Near such points, the dynamics is often effectively restricted to a
two-dimensional subspace, leading to a universal two-level structure that governs
the local spectral behavior. This mechanism underlies the Landau--Zener
description of parametric level crossings~\cite{Landau1932, Zener1932,
Shevchenko2010} and provides a natural setting in which information-geometric
quantities can be computed exactly. From a statistical-mechanical perspective,
the integrated geometric susceptibility associated with such crossings encodes
the total distinguishability accumulated across the parameter space, providing
a compact measure of the system's collective response to external field variations.

Two-dimensional Dirac fermions under Aharonov--Bohm flux provide a
particularly transparent realization of this mechanism. The flux shifts the
effective angular momentum and, at integer values, compensates a given angular
sector, enhancing spectral sensitivity and enabling zero-mode formation in the
chiral limit~\cite{Jackiw1984, Gerbert1989}. This chiral limit plays a role
analogous to a critical point in statistical mechanics: the vanishing of the
mass gap signals a qualitative reorganization of the low-energy spectrum,
accompanied by a divergence of the geometric susceptibility that parallels
the divergence of thermodynamic susceptibilities at continuous phase transitions.
Information-theoretic aspects of Dirac--Aharonov--Bohm systems, including
entropy-based measures and spectral sensitivity analyses, have been investigated
in various contexts~\cite{Lima2023, Edet2022}.

In this work, we analyze the Bures metric associated with flux variations in
two-dimensional Dirac systems and establish its universal structure near integer
values of the reduced flux. We show that the leading flux-dependent contribution
to the ground-state fidelity response is governed by an effective two-level
Hamiltonian exhibiting an avoided crossing controlled by the Dirac mass. Within
this framework, the Bures metric can be computed exactly and displays a
Lorentzian-type peak centered at integer flux values. We further introduce an
integrated geometric susceptibility that diverges as $\chi(m) \sim 1/m$ in the
chiral limit, exhibiting inverse-mass scaling consistent with the
statistical-mechanical analogy of gap-closing criticality.

A central point emphasized in this work is that the geometric response induced
by flux insertion in Dirac systems is not tied to Berry curvature, topology, or
phase transitions in the conventional sense, but instead emerges from a purely
local spectral mechanism associated with the compensation of angular momentum
at integer values of the Aharonov--Bohm flux. The resulting Bures metric
therefore provides a direct geometric characterization of flux-induced spectral
rearrangements, placing the analysis on a different conceptual footing from
standard applications of fidelity susceptibility while retaining its
statistical-mechanical interpretation as a measure of parametric sensitivity.
This highlights information geometry as a compact and physically meaningful
language to encode susceptibility-type responses in gauge-controlled relativistic
systems.

Finally, we show that the Lorentzian profile of the metric admits a simple
geometric interpretation in terms of the Fubini--Study metric on the Bloch
sphere~\cite{Anandan1990, Pati1991}, making explicit the relation between
flux-induced spectral rearrangements and the curvature of the ground-state
manifold.

The paper is organized as follows. In Sect.~2 we review the Dirac Hamiltonian
under Aharonov--Bohm flux insertion and discuss the spectral structure near
integer values of the reduced flux parameter. Sect.~3 introduces the effective
two-level description capturing the leading flux dependence; the microscopic
derivation of this effective Hamiltonian through a controlled projection of the
full Dirac--Aharonov--Bohm operator is presented in Sect.~8, where it is shown
that the effective parameters $E_0$ and $a$ follow directly from the radial
eigenfunctions. In Sect.~4 we compute the Bures metric exactly within the
effective model. Sect.~5 provides its physical interpretation for
Dirac--Aharonov--Bohm systems. Sect.~6 analyzes the integrated geometric
susceptibility $\chi(m) = \pi/(8m)$ and its inverse-mass scaling behavior.
Sect.~7 presents the geometric interpretation in terms of the Bloch-sphere
structure of the ground-state manifold. Sect.~9 contains our discussion and
conclusions.

\section{Dirac fermions under flux insertion: spectral structure}

We consider a two-dimensional massive Dirac fermion subjected to an
Aharonov--Bohm flux localized at the origin. In polar coordinates, the
Hamiltonian can be written as
\begin{equation}
H = -i\sigma^r \partial_r - \frac{i}{r}\sigma^\phi(\partial_\phi - i\lambda)
+ m\sigma^z,
\label{eq:Hfull}
\end{equation}
where $\lambda$ denotes the reduced magnetic flux in units of the flux quantum,
and $\sigma^r$, $\sigma^\phi$ are the radial and angular Pauli matrices.

Using the standard angular decomposition
\begin{equation}
\psi(r,\phi) = e^{i\ell\phi}
\begin{pmatrix} f_\ell(r) \\ g_\ell(r)e^{i\phi} \end{pmatrix},
\label{eq:decomp}
\end{equation}
one finds that the flux enters through the effective angular momentum index
\begin{equation}
\nu(\lambda) = \ell - \lambda.
\label{eq:nu}
\end{equation}
The radial equations depend parametrically on $\nu$, and the spectrum is
therefore continuously deformed as the flux is varied. In particular, integer
values of $\lambda$ correspond to points where the effective angular momentum
vanishes in a given sector. Near such points, the spectral structure simplifies
and the lowest-energy states in adjacent angular momentum sectors become nearly
degenerate.

From a statistical-mechanical perspective, this continuous deformation of the
spectrum under the control parameter $\lambda$ is directly analogous to the
tuning of an external field in a thermodynamic system. The enhanced sensitivity
at integer flux values plays the role of a pseudocritical point, where the
ground-state response function --- here, the Bures metric --- reaches its
maximum value.

The detailed spectrum depends on the infrared regularization (e.g., boundary
conditions in a finite disk or momentum cutoffs in the infinite plane). However,
independently of these details, the flux dependence near integer values of
$\lambda$ is governed by the behavior of $\nu(\lambda)$ close to zero. In this
regime, the low-energy dynamics is effectively restricted to a two-dimensional
subspace spanned by the states whose energies are most sensitive to variations
of $\nu$.

This observation motivates the introduction of an effective two-level description
valid in the vicinity of integer flux values. In the next section, we construct
this effective Hamiltonian and analyze the associated Bures metric exactly.

\section{Effective two-level description near integer flux}

As discussed in the previous section, integer values of the reduced flux
parameter $\lambda$ correspond to points where the effective angular momentum
index $\nu(\lambda) = \ell - \lambda$ vanishes in a given angular momentum
sector. In the vicinity of such points, the low-energy dynamics is dominated by
states whose energies are most sensitive to variations of $\nu$.

Independently of the specific infrared regularization, the Dirac Hamiltonian near
$\nu = 0$ exhibits a local structure that can be captured by an effective
two-dimensional subspace. This reduction reflects the fact that the flux
dependence enters linearly through $\nu(\lambda)$ in the angular part of the
operator, while the mass term couples the spinor components.

Motivated by this structure, we introduce an effective two-level Hamiltonian of
the form
\begin{equation}
H_{\rm eff}(\nu) = \nu\,\sigma_x + m\,\sigma_z,
\label{eq:Heff}
\end{equation}
where $\nu = \ell - \lambda$ measures the deviation from the integer-flux point
and $m$ is the Dirac mass. This Hamiltonian encodes the leading dependence of
the low-energy sector on the control parameter $\lambda$.

The spectrum of Eq.~\eqref{eq:Heff} is given exactly by
\begin{equation}
E_\pm(\nu) = \pm\sqrt{m^2 + \nu^2}.
\label{eq:spectrum}
\end{equation}
Equation~\eqref{eq:spectrum} describes an avoided level crossing controlled by
the mass $m$. In the chiral limit $m \to 0$, the gap closes linearly in $\nu$,
signaling the onset of zero-mode formation. For finite $m$, the minimal gap at
$\nu = 0$ is $2m$.

The structure of Eq.~\eqref{eq:Heff} is formally identical to a spin-$1/2$
particle in a two-component effective field $(\nu, 0, m)$, a model widely
studied in the context of quantum statistical mechanics and quantum information. The microscopic justification of this effective Hamiltonian -- including
the explicit derivation of $E_0$ and $a$ from the radial eigenfunctions of
the full Dirac-Aharonov-Bohm operator -- is presented in Sect.~8. That
analysis shows that both parameters are insensitive, at leading order, to
the choice of infrared regularization, confirming the universality of the
effective model. The reader primarily interested in the physical results
may read Sects.~3--7 independently, treating the effective Hamiltonian as
an input, before consulting the derivation.

This analogy makes explicit the connection between the flux-tuned spectral
rearrangement and the response theory of two-level systems, and provides the
foundation for an exact computation of the Bures metric in the following section. 
\section{Bures metric in the effective two-level model}

We now compute the Bures metric associated with variations of the reduced flux
parameter $\lambda$ within the effective two-level description introduced in the
previous section.

Since $\nu = \ell - \lambda$, variations of $\lambda$ are equivalent to
variations of $\nu$ up to a sign, and the effective Hamiltonian reads
\begin{equation}
H_{\rm eff}(\nu) = \nu\,\sigma_x + m\,\sigma_z.
\end{equation}
As shown in Sect.~8, the microscopic projection of the Dirac--Aharonov--Bohm
operator onto the low-energy subspace yields a Hamiltonian of the general form
$a\nu\,\sigma_x + E_0\,\sigma_z$, where $E_0$ is the lowest eigenvalue at
integer flux and $a$ is the flux-induced coupling matrix element. In the
universal regime $mR \gg 1$, one has $E_0 \to m$ and the coupling $a$
introduces an overall rescaling that does not affect the Lorentzian structure
of the metric. In what follows, we work in units where $E_0 = m$ and $a = 1$,
which captures the universal behavior and simplifies the presentation.

The normalized ground state $|{-}\rangle$ of this Hamiltonian corresponds to
the negative-energy eigenvalue
\begin{equation}
E_-(\nu) = -\sqrt{m^2 + \nu^2}.
\end{equation}
For a two-level system depending on a single real parameter, the Bures metric
(or fidelity susceptibility) admits the spectral representation
\begin{equation}
g_{\lambda\lambda} = \frac{|\langle +|\partial_\lambda H_{\rm eff}|-\rangle|^2}
{(E_+ - E_-)^2},
\label{eq:gspec}
\end{equation}
where $|{+}\rangle$ denotes the excited state and
$E_+ - E_- = 2\sqrt{m^2+\nu^2}$.

Since $\partial_\lambda = -\partial_\nu$, one has
\begin{equation}
\partial_\lambda H_{\rm eff} = -\sigma_x.
\end{equation}
A straightforward computation yields
\begin{equation}
|\langle +|\sigma_x|-\rangle|^2 = \frac{m^2}{m^2+\nu^2}.
\end{equation}
Substituting into Eq.~\eqref{eq:gspec}, we obtain
\begin{equation}
g_{\lambda\lambda} = \frac{1}{4}\,\frac{m^2}{(m^2+\nu^2)^2}.
\label{eq:metric}
\end{equation}

\begin{figure}[t]
  \centering
  \includegraphics[width=\linewidth]{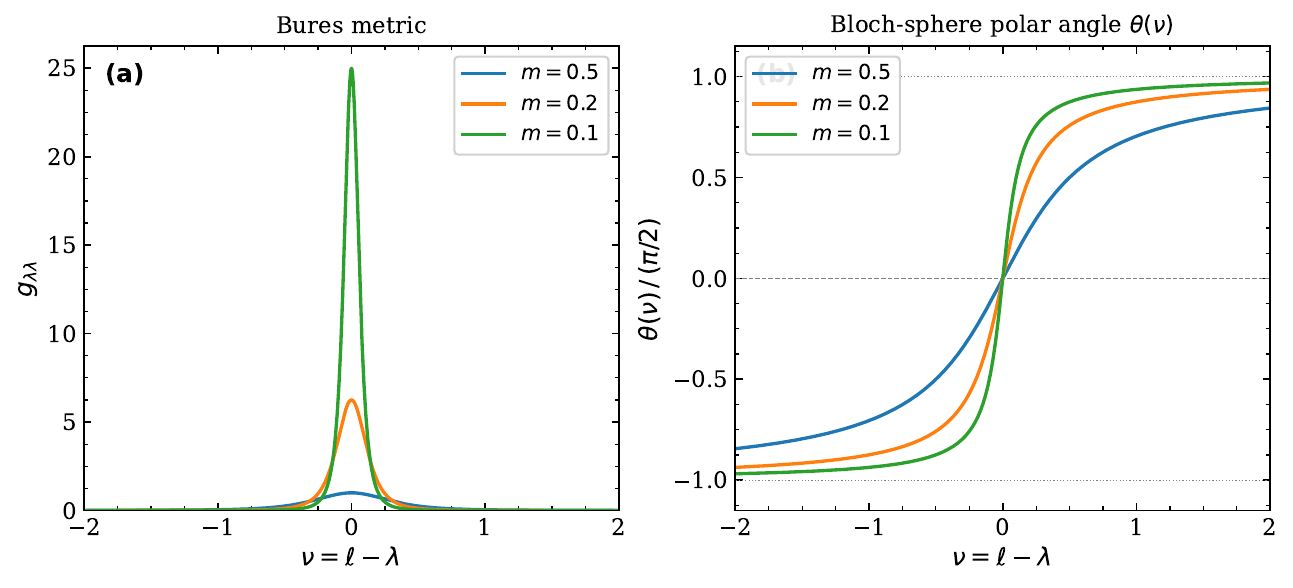}
  \caption{%
    \textbf{(a)}~Bures metric $g_{\lambda\lambda}(\nu)$ in the
    effective two-level model for $m = 0.5$, $0.2$, and $0.1$.
    The Lorentzian peak at $\nu = 0$ scales as
    $g_{\lambda\lambda}^{\max} \propto m^{-2}$ and diverges in the
    chiral limit.
    \textbf{(b)}~Normalized Bloch-sphere polar angle
    $(2/\pi)\arctan(\nu/m)$ for the same masses.
    Since $g_{\lambda\lambda} = \tfrac{1}{4}(d\theta/d\nu)^2$
    (Sect.~7), the curvature of the ground-state trajectory on the
    Bloch sphere directly generates the Lorentzian profile in
    panel~(a).}
  \label{fig:bures_bloch}
\label{fig:bures_combined}
\end{figure}

The behavior of Eq.~(\ref{eq:metric}) is illustrated in Fig.~\ref{fig:bures_bloch}(a).
The geometric origin of the Lorentzian profile is made explicit in
Fig.~\ref{fig:bures_bloch}(b), which shows the Bloch-sphere polar angle
$\theta(\nu) = \arctan(\nu/m)$: since $g_{\lambda\lambda} =
\tfrac{1}{4}(d\theta/d\nu)^2$, the metric peak at $\nu = 0$ directly
reflects the maximum curvature of the ground-state trajectory on the
Bloch sphere (Sect.~7).

The metric exhibits a pronounced peak at $\nu=0$,
corresponding to the avoided crossing point.
As the mass parameter decreases, the peak becomes
narrower and higher, reflecting the enhanced
parametric sensitivity in the vicinity of gap closing.

Note that Eq.~\eqref{eq:metric} is exact within the regime where the two-level
reduction remains valid. The Bures metric exhibits a Lorentzian-type peak
centered at $\nu = 0$, with a width controlled by the mass $m$. In the
language of statistical mechanics, this profile is directly analogous to a
Lorentzian susceptibility peak in a driven two-level system, where $m$ plays
the role of a gap-controlling field and $\nu$ measures the distance from
resonance.

The maximal value occurs at integer flux ($\nu = 0$),
\begin{equation}
g^{\rm max}_{\lambda\lambda} = \frac{1}{4m^2}.
\label{eq:gmax}
\end{equation}
In the chiral limit $m \to 0$, the peak becomes increasingly sharp and diverges
as $1/m^2$, reflecting the approach to a gap-closing point. The characteristic
dependence on $(m^2 + \nu^2)$ is a direct consequence of the two-level structure
encoded in Eq.~\eqref{eq:Heff}.

\section{Physical interpretation for Dirac--Aharonov--Bohm systems}
In the Dirac--Aharonov--Bohm Hamiltonian, the flux enters solely through
the effective angular momentum index $\nu(\lambda) = \ell - \lambda$. Near
integer flux, the spectral response is therefore governed by the states
most sensitive to $\nu$, leading to an effectively two-dimensional
low-energy structure.

Although the complete spectrum depends on the chosen infrared
regularization (e.g., boundary conditions in a finite disk or momentum
cutoffs in the infinite plane), the local structure near integer flux
values is insensitive to these details at leading order. In this regime,
the relevant part of the Hilbert space is effectively two-dimensional and
can be described by the Hamiltonian $H_{\rm eff}(\nu) = \nu\sigma_x +
    m\sigma_z$.

Within this effective description, the Bures metric acquires the form
given in Eq.~(11). This expression captures the enhanced parametric
sensitivity of the ground state as the spectral gap approaches its minimum
at integer flux. The peak of the metric at $\nu = 0$ reflects the
increased distinguishability between nearby quantum states when the system
is close to a gap-closing configuration --- a behavior that is the
quantum-geometric analogue of the enhanced fluctuations observed near thermodynamic critical points.

\paragraph{Connection to persistent currents and orbital magnetization.}

The physical content of the Bures metric in this setting goes beyond a
formal analogy. In the Dirac--Aharonov--Bohm Hamiltonian, variations of
the reduced flux $\lambda$ enter through the gauge field, and the operator
$\partial_\lambda H$ is directly proportional to the azimuthal current
density $J_\varphi$ circulating around the flux tube. Within the effective
two-level description, this gives
\begin{equation}
  \partial_\lambda H_{\rm eff} = -\sigma_x,
\end{equation}
and the spectral representation of the Bures metric (Eq.~(8)) becomes
\begin{equation}
  g_{\lambda\lambda}
  = \frac{|\langle +|\,\partial_\lambda H_{\rm eff}\,|-\rangle|^2}
         {(E_+ - E_-)^2}
  = \frac{|\langle +|\sigma_x|-\rangle|^2}{4(m^2+\nu^2)},
\end{equation}
which is precisely the zero-frequency limit of the current--current
response function projected onto the low-energy subspace. This identifies
$g_{\lambda\lambda}$ as the \emph{geometric (paramagnetic) contribution to
the persistent current susceptibility}. More explicitly, the total
persistent current susceptibility decomposes as
\begin{equation}
  \chi_I
  = -\frac{\partial^2 E_-}{\partial\lambda^2} + 2\,g_{\lambda\lambda},
  \label{eq:chi_decomp}
\end{equation}
where the first term is the diamagnetic contribution and
$2g_{\lambda\lambda}$ is the paramagnetic (geometric) part directly given
by the Bures metric. Near integer flux ($\nu \approx 0$), the geometric
contribution dominates and reaches its maximum value
$g_{\lambda\lambda}^{\max} = 1/(4m^2)$.

The two contributions are compared quantitatively in
Fig.~\ref{fig:chi_decomp}(b) as functions of $m$; the geometric term dominates for   $m < 1/2$, as shown by the shaded region.

To make this connection quantitative, consider a mesoscopic graphene ring
on a hexagonal boron nitride (hBN) substrate, where the substrate-induced
gap is $m \sim 1\,\mathrm{meV}$~\cite{Giovannetti2007} and a ring radius
$R \sim 1\,\mu\mathrm{m}$ is experimentally accessible~\cite{Russo2008}.
The flux-induced coupling evaluates to $a \sim \hbar v_F/R \sim
0.6\,\mathrm{meV}$, placing the system well within the regime $|\nu| \ll 1$
where the effective description is quantitatively reliable. The peak
geometric susceptibility $g_{\lambda\lambda}^{\max} = 1/(4m^2)$
corresponds to a persistent current susceptibility of order $(e/h)^2/m$,
which for $m \sim 1\,\mathrm{meV}$ yields a signal consistent with the
sensitivity of existing Aharonov--Bohm interferometry experiments in
graphene rings~\cite{Russo2008}.

In the chiral limit $m \to 0$, the effective gap vanishes and the Bures
metric diverges, indicating the singular nature of the zero-mode formation
induced by flux insertion. For finite mass, the gap remains nonzero and
the metric peak is smoothed accordingly.

It is important to emphasize that this behavior does not rely on Berry
curvature. Since the parameter space is one-dimensional, the Berry
curvature vanishes identically, whereas the Bures metric remains sensitive
to gap variations. In this sense, the information-geometric response
provides a complementary probe of flux-induced spectral rearrangements in
Dirac systems, with the added advantage of being directly linked to a
measurable mesoscopic observable --- the persistent current susceptibility
--- through the decomposition~(\ref{eq:chi_decomp}).

\section{Integrated Bures response and scaling behavior}

The Lorentzian-type peak of the Bures metric obtained in Eq.~\eqref{eq:metric}
suggests introducing an integrated susceptibility that quantifies the total
geometric response associated with variations of the control parameter. This
quantity is the natural information-geometric analogue of the integrated
susceptibility familiar from statistical mechanics, which measures the total
response of a thermodynamic system to an external perturbation integrated
over the full range of the driving parameter.

Before proceeding, we note that the integration over the entire real axis
in Eq.~(16) is to be understood within the effective theory, where $\nu =
\ell - \lambda$ is treated as a continuous variable in the vicinity of a
single avoided crossing. The physical Aharonov--Bohm flux is periodic with
period~1 (in units of the flux quantum), so the full physical response
accumulates contributions from all integer-flux crossing points. The
quantity $\chi(m)$ therefore represents the integrated geometric response
associated with a \emph{single} such avoided crossing in the effective
theory --- a well-defined and physically meaningful quantity analogous to
the integrated susceptibility of an isolated resonance in a driven
two-level system. The extension of the integration to $\pm\infty$
introduces only exponentially small errors beyond the regime of validity
of the effective description, given that $g_{\lambda\lambda}$ decays as
$\nu^{-4}$ for $|\nu| \gg m$, rendering the tail contributions negligible.

\begin{figure}[t]
  \centering
  \includegraphics[width=\linewidth]{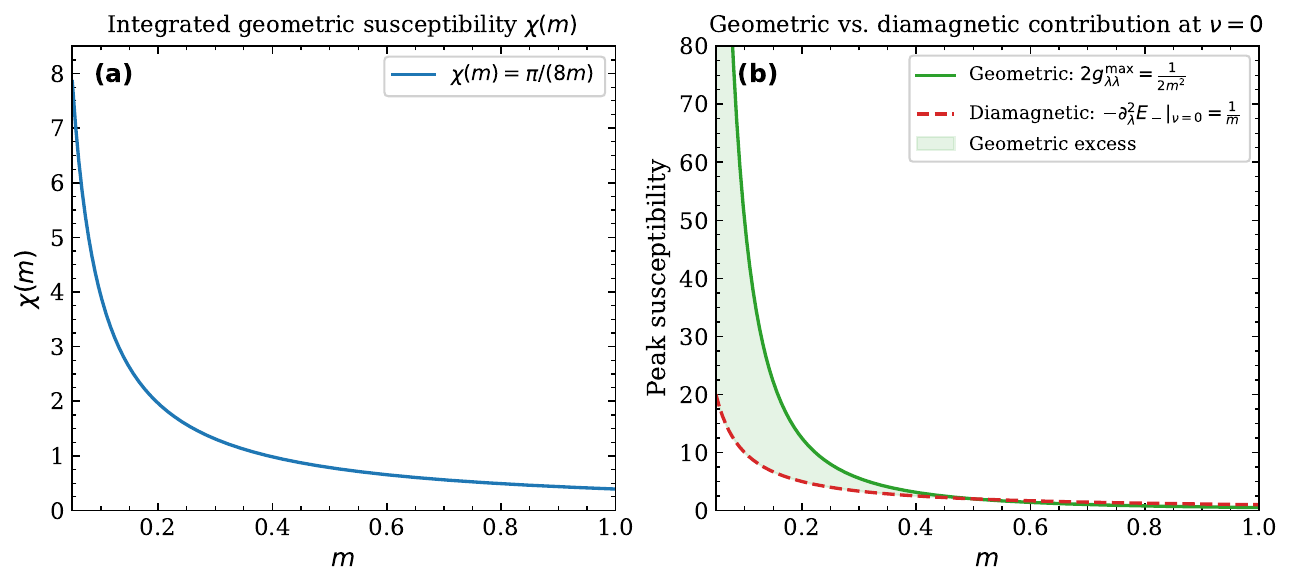}
  \caption{%
    \textbf{(a)}~Integrated Bures response $\chi(m)=\pi/(8m)$,
    representing the total geometric response of a single avoided
    crossing in the effective theory (Sect.~6).
    \textbf{(b)}~Geometric (paramagnetic) contribution
    $2g_{\lambda\lambda}^{\max}=1/(2m^2)$ and diamagnetic
    contribution $-\partial^2_\lambda E_-\big|_{\nu=0}=1/m$ to the
    peak persistent current susceptibility at $\nu=0$, as functions
    of $m$. The shaded region marks the geometric excess; for
    $m < 1/2$ the Bures metric dominates the total susceptibility.}
  \label{fig:chi_decomp}
\end{figure}

For the effective two-level model, we define
\begin{equation}
\chi(m) = \int_{-\infty}^{+\infty} g_{\lambda\lambda}\,d\lambda.
\label{eq:chi}
\end{equation}
Since $\nu = \ell - \lambda$, one has $d\lambda = -d\nu$, and using
Eq.~\eqref{eq:metric} we obtain
\begin{equation}
\chi(m) = \frac{1}{4}\int_{-\infty}^{+\infty}
\frac{m^2}{(m^2+\nu^2)^2}\,d\nu.
\label{eq:chiint}
\end{equation}
The integral can be evaluated exactly,
\begin{equation}
\int_{-\infty}^{+\infty}\frac{m^2}{(m^2+\nu^2)^2}\,d\nu = \frac{\pi}{2m},
\end{equation}
which yields
\begin{equation}
\chi(m) = \frac{\pi}{8m}.
\label{eq:chiresult}
\end{equation}
The scaling behavior of Eq.~\eqref{eq:chiresult} is shown in Fig.~\ref{fig:chi_decomp}(a).

The inverse dependence on the mass parameter becomes increasingly pronounced as the system approaches the chiral limit.


This inverse-mass scaling reflects the enhanced parametric sensitivity near
gap closing, consistent with fidelity susceptibilities at avoided
crossings~\cite{Zanardi2006, Gu2010}. The behavior $\chi(m) \sim m^{-1}$
is the direct analogue, in the present information-geometric context, of the
power-law divergence of thermodynamic susceptibilities near a critical point,
with the Dirac mass $m$ playing the role of a relevant coupling that controls
the distance from criticality.

In the present context, the divergence of $\chi(m)$ signals that the total
distinguishability accumulated across the avoided crossing becomes increasingly
large as the mass decreases, consistent with the approach to zero-mode formation
in the Dirac system. This result provides a quantitative, analytically exact
characterization of the statistical-mechanical response of a relativistic quantum
system to gauge-field insertion, and may serve as a benchmark for more complex
interacting or disordered Dirac systems where exact results are unavailable.

\section{Geometric structure of the ground-state manifold}

The effective two-level Hamiltonian $H_{\rm eff}(\nu) = \nu\sigma_x + m\sigma_z$
admits a simple geometric interpretation in terms of the Bloch sphere
representation of pure states. This interpretation connects the flux-induced
susceptibility directly to the curvature of the ground-state manifold in
projective Hilbert space, providing a purely geometric --- and
parameter-independent --- explanation for the Lorentzian profile of the Bures
metric.

The ground state can be written as
\begin{equation}
|{-}\rangle = \frac{1}{\sqrt{2E(E+m)}}
\begin{pmatrix} E+m \\ \nu \end{pmatrix},
\qquad E = \sqrt{m^2+\nu^2}.
\label{eq:gs}
\end{equation}
The Fubini--Study metric on the Bloch sphere induces the line element~\cite{Anandan1990, Pati1991}
\begin{equation}
ds^2 = \frac{1}{4}\,d\theta^2.
\label{eq:FS}
\end{equation}
Using
\begin{equation}
\frac{d\theta}{d\nu} = \frac{m}{m^2+\nu^2},
\label{eq:dtheta}
\end{equation}
we recover
\begin{equation}
ds^2 = \frac{1}{4}\,\frac{m^2}{(m^2+\nu^2)^2}\,d\nu^2,
\label{eq:dsBloch}
\end{equation}
which coincides exactly with the Bures metric derived in Eq.~\eqref{eq:metric}. This geometric picture is illustrated in Fig.~\ref{fig:bures_bloch}(b):
the normalized angle $\theta(\nu)/(\pi/2)$ rotates most rapidly near
$\nu = 0$, where the ground state undergoes maximal reorganization on
the Bloch sphere, and this rapid rotation is precisely what produces
the Lorentzian peak in $g_{\lambda\lambda}$.

The Lorentzian profile therefore reflects the curvature of the projective Hilbert
space along the flux-induced path. No additional dynamical ingredient is required:
the enhancement of geometric susceptibility near integer flux values is entirely
encoded in the geometry of the ground-state manifold. This result establishes
a direct bridge between the statistical-mechanical susceptibility $\chi(m)$
introduced in the previous section and the differential geometry of quantum
state space, showing that the divergence of the integrated response in the chiral
limit is a purely geometric consequence of the increasing curvature of the
ground-state trajectory on the Bloch sphere as $m \to 0$.

\section{Controlled low-energy projection of the Dirac--Aharonov--Bohm
Hamiltonian}

We now establish explicitly the connection between the Dirac--Aharonov--Bohm
Hamiltonian and the effective two-level description introduced previously. The
derivation proceeds by projecting the full operator onto the low-energy subspace
near integer values of the reduced flux parameter.

\subsection{Radial Hamiltonian and flux dependence}

After angular decomposition,
\begin{equation}
\psi(r,\phi) = e^{i\ell\phi}
\begin{pmatrix} f(r) \\ g(r)e^{i\phi} \end{pmatrix},
\end{equation}
the Dirac Hamiltonian in the presence of Aharonov--Bohm flux reduces to a radial
operator
\begin{equation}
H_\ell(\lambda) =
\begin{pmatrix} m & D_\nu \\ D_\nu^\dagger & -m \end{pmatrix},
\label{eq:Hradial}
\end{equation}
where
\begin{equation}
D_\nu = -i\!\left(\partial_r + \frac{\nu+1}{r}\right), \qquad \nu = \ell - \lambda.
\end{equation}
The flux dependence enters exclusively through the effective angular momentum
index $\nu$. Expanding around $\nu = 0$,
\begin{equation}
D_\nu = D_0 + \nu V, \qquad V = -\frac{i}{r}.
\end{equation}
Accordingly, the radial Hamiltonian admits the exact decomposition
\begin{equation}
H_\ell(\lambda) = H_0 + \nu W, \qquad
W = \begin{pmatrix} 0 & V \\ V^\dagger & 0 \end{pmatrix},
\label{eq:Hdecomp}
\end{equation}
where $H_0 \equiv H_\ell(\lambda)\big|_{\nu=0}$ denotes the radial Dirac
Hamiltonian evaluated at integer flux,
\begin{equation}
H_0 = \begin{pmatrix} m & D_0 \\ D_0^\dagger & -m \end{pmatrix}.
\end{equation}

\subsection{Projection onto the low-energy subspace}

Let $|u_\pm\rangle$ denote the two eigenstates of $H_0$ with lowest positive
and negative energies,
\begin{equation}
H_0|u_\pm\rangle = \pm E_0|u_\pm\rangle.
\end{equation}
We define the projector
\begin{equation}
P = |u_+\rangle\langle u_+| + |u_-\rangle\langle u_-|.
\end{equation}
Projecting Eq.~\eqref{eq:Hdecomp} onto this subspace yields
\begin{equation}
H_{\rm eff}(\nu) = PH_\ell(\lambda)P = PH_0P + \nu PWP.
\end{equation}
In the basis $\{|u_+\rangle, |u_-\rangle\}$, one has $PH_0P = E_0\sigma_z$.
The flux-induced coupling is determined by the matrix element
\begin{equation}
a = \langle u_+|W|u_-\rangle
= \int_0^R dr\,r\left[f_+^*(r)\!\left(-\frac{i}{r}\right)g_-(r)
+ g_+^*(r)\!\left(\frac{i}{r}\right)f_-(r)\right],
\end{equation}
where $(f_\pm, g_\pm)$ denote the radial spinor components of $|u_\pm\rangle$.
Hermiticity implies that $a$ is real and that diagonal matrix elements vanish
by symmetry, so that the projected Hamiltonian takes the form
\begin{equation}
H_{\rm eff}(\nu) = a\nu\,\sigma_x + E_0\,\sigma_z.
\label{eq:Heffproj}
\end{equation}

\subsection{Effective spectrum and information geometry}

The spectrum of Eq.~\eqref{eq:Heffproj} is
\begin{equation}
E_\pm(\nu) = \pm\sqrt{E_0^2 + a^2\nu^2}.
\end{equation}
Within this controlled low-energy projection, the Bures metric associated with
variations of $\lambda$ is obtained exactly as
\begin{equation}
g_{\lambda\lambda} = \frac{1}{4}\,\frac{a^2 E_0^2}{(E_0^2 + a^2\nu^2)^2}.
\label{eq:gproj}
\end{equation}

The nontrivial aspect in the present context is that the effective
two-level structure emerges from a controlled projection of the
Dirac--Aharonov--Bohm operator, rather than being postulated
phenomenologically.

Equation~(33) shows that the Lorentzian-type structure of the metric
arises directly from the projected Dirac Hamiltonian. The coefficients
$E_0$ and $a$ are determined by the radial eigenfunctions of the original
Dirac--Aharonov--Bohm problem and therefore encode the dependence on the
chosen infrared regularization. The universality of the Lorentzian
structure --- i.e., the fact that at leading order in $1/(mR)^2$ the
profile reduces to
\begin{equation}
  g_{\lambda\lambda}
  = \frac{a^2 E_0^2}{(E_0^2 + a^2\nu^2)^2}
  \;\xrightarrow{\;mR\gg 1\;}\;
  \frac{m^2}{(m^2+\nu^2)^2}
  + \mathcal{O}\!\left(\frac{1}{(mR)^2}\right),
\end{equation}
independently of the specific regularization --- is a nontrivial output of
the projection. It follows from the parametric suppression of finite-size
corrections, as shown explicitly in Sect.~8.4, rather than from any
additional assumption about the system.

The validity of this reduction requires $|\nu|$ to remain sufficiently
small so that couplings to higher-energy states are suppressed by energy
denominators of order $|E_n - E_0| \gg |a\nu|$. Within this regime, the
effective Hamiltonian~(31) provides a quantitatively controlled description
of the flux-induced low-energy response.

\subsection{Explicit scaling of $E_0$ and $a$ for a finite disk}

To make explicit the controlled nature of the low-energy projection
leading to the effective Hamiltonian~(31), we analyze the scaling of the
coefficients $E_0$ and $a$ for a concrete infrared regularization. We
consider a finite disk of radius $R$ and impose MIT bag boundary
conditions at the edge,
\begin{equation}
  \bigl(1 + i\sigma_r\bigr)\psi(r,\varphi)\big|_{r=R} = 0,
\end{equation}
which provide a standard confinement for two-dimensional Dirac fermions
and avoid spurious edge modes.

\paragraph{Lowest-energy scale at integer flux.}
At integer values of the reduced flux parameter, corresponding to $\nu =
\ell - \lambda = 0$, the radial Hamiltonian $H_0$ has a discrete spectrum.
The positive-energy eigenvalues take the generic form
\begin{equation}
  E_n^2 = m^2 + k_n^2,
\end{equation}
where the radial momenta $k_n$ are quantized by the boundary condition and
scale as $k_n \sim \mathcal{O}(1/R)$. The lowest positive eigenvalue
therefore satisfies
\begin{equation}
  E_0 = \sqrt{m^2 + k_1^2} = m + \mathcal{O}\!\left(\frac{1}{mR^2}\right),
  \qquad mR \gg 1.
\end{equation}
In this regime, the energy of the projected doublet is set predominantly
by the Dirac mass, while finite-size corrections are parametrically
suppressed by $1/(mR)^2$.

\paragraph{Scaling of the flux-induced coupling.}
The off-diagonal coefficient $a$ is defined by the matrix element $a =
\langle u_+|W|u_-\rangle$, with $W$ containing the operator $1/r$. The
scaling of $a$ follows from dimensional analysis combined with
wave-function normalization. The low-energy radial eigenfunctions $|u_\pm
\rangle$ are smooth over the disk and normalized within an area of order
$R^2$, implying typical amplitudes $u_\pm(r) \sim R^{-1}$.
Consequently,
\begin{equation}
  a \sim \int_0^R dr\, r\, \frac{1}{r}\, u_+(r)\,u_-(r)
  \sim \mathcal{O}\!\left(\frac{1}{R}\right).
\end{equation}
This scaling is robust and independent of microscopic details of the
confinement, depending only on the finite size of the system.

\paragraph{Universal form of the Lorentzian profile.}
With $E_0 = m + \mathcal{O}(1/(mR^2))$ and $a \sim \mathcal{O}(1/R)$, the
Bures metric of Eq.~(33) satisfies
\begin{equation}
  g_{\lambda\lambda}
  = \frac{a^2 E_0^2}{\bigl(E_0^2 + a^2\nu^2\bigr)^2}
  \;\xrightarrow{\;mR\gg 1\;}\;
  \frac{m^2}{(m^2 + \nu^2)^2}
  + \mathcal{O}\!\left(\frac{1}{(mR)^2}\right).
\end{equation}
The Lorentzian structure of the metric is therefore insensitive to the
choice of infrared regularization at leading order in $1/(mR)^2$. While
this derivation has been carried out for MIT bag boundary conditions, the
conclusion is expected to hold for any regularization preserving the
rotational symmetry of the problem, since the scaling of $E_0$ and $a$
follows from symmetry and dimensional analysis rather than from the
specific boundary conditions.

\paragraph{Validity of the two-level reduction.}
At $\nu = 0$, the next radial excitation is separated from the projected
doublet by a gap $\Delta = E_2 - E_0 \sim \mathcal{O}(1/R)$. The
effective Hamiltonian therefore provides a controlled description provided
$|a\nu| \ll \Delta$, which translates into the condition $|\nu| \ll 1$.
This is precisely the regime of flux values close to an integer, where the
avoided-crossing structure dominates the spectral response.

Taken together, these results show explicitly that the parameters entering
the effective two-level Hamiltonian are not phenomenological but follow
from a controlled projection of the Dirac--Aharonov--Bohm problem, and
that the universality of the Lorentzian profile is a nontrivial consequence
of this projection rather than an additional assumption.

\section{Conclusions}
Within the effective two-level description that governs the low-energy
response near integer values of the reduced flux, we have shown that
magnetic flux insertion in two-dimensional Dirac systems induces a
universal geometric response governed by an avoided crossing. Through a
controlled low-energy projection of the Dirac-Aharonov-Bohm Hamiltonian,
we derived an exact expression for the Bures metric within this effective
theory and established its Lorentzian structure near integer flux values.

The resulting geometric susceptibility provides a direct measure of
flux-induced spectral rearrangements without invoking Berry curvature or
topological transitions. Within the information-geometric framework
developed here, the Bures metric plays the role of a quantum response
function analogous to thermodynamic susceptibilities in classical
statistical mechanics: it quantifies the sensitivity of the ground-state
structure to external perturbations and diverges when the spectral gap
closes. The integrated geometric susceptibility $\chi(m) = \pi/(8m)$
exhibits inverse-mass scaling that is the direct information-geometric
counterpart of power-law divergences at critical points, with the Dirac
mass acting as a relevant coupling controlling the distance from the chiral
fixed point.

An important result of the present work is the identification of
$g_{\lambda\lambda}$ as the geometric (paramagnetic) contribution to the
persistent current susceptibility, as expressed in Eq.~(\ref{eq:chi_decomp}).
This connection elevates the Bures metric from a formal information-geometric
quantity to a physically measurable response function in mesoscopic Dirac
systems, and provides a concrete prediction for Aharonov--Bohm
interferometry experiments.

These results place the analysis on a clear statistical-mechanical footing:
the flux-tuned Dirac system provides an analytically tractable model in
which the interplay between gauge fields, quantum geometry, and
susceptibility divergences can be worked out exactly within the projected
theory, without relying on semiclassical approximations or perturbation
theory. The exact result $\chi(m) = \pi/(8m)$ constitutes a rare
closed-form expression for an integrated geometric susceptibility in a
relativistic quantum system, and may serve as a benchmark for more complex
interacting or disordered Dirac systems where exact results are
unavailable. The framework developed here can be extended to
multi-parameter deformations, interacting systems, and curved backgrounds,
where richer geometric structures may arise.

For realistic Dirac materials such as graphene on hBN
substrates~\cite{Giovannetti2007} or Kekul\'e-distorted
lattices~\cite{Gomes2012}, induced gaps typically lie in the meV range,
corresponding to mass parameters $m$ for which the predicted Lorentzian
peak $g_{\lambda\lambda} = m^2/[4(m^2 + \nu^2)^2]$ is narrow and strongly
enhanced. In mesoscopic graphene rings, flux variations corresponding to a
single flux quantum are experimentally accessible~\cite{Russo2008}, placing
the relevant range of $\nu$ well within the regime where the Lorentzian
profile dominates. For $m \sim 1\,\mathrm{meV}$ and $R \sim 1\,\mu\mathrm{m}$,
the peak geometric susceptibility $g_{\lambda\lambda}^{\max} = 1/(4m^2)$
corresponds to a persistent current susceptibility of order $(e/h)^2/m$,
which is in principle detectable in existing Aharonov--Bohm interferometry
setups. The integrated susceptibility $\chi(m) = \pi/(8m)$ provides a
direct quantitative prediction for the total geometric response accumulated
over a single flux period, which could be extracted from measurements of
persistent current susceptibility or orbital magnetization as a function of
applied flux. These considerations suggest that the predicted geometric
response may be observable in existing experimental setups, providing a
possible route toward probing information-geometric susceptibilities in
Dirac systems.

An important result of the present work is the identification of
$g_{\lambda\lambda}$ as the geometric (paramagnetic) contribution to the
persistent current susceptibility, as expressed in Eq.~(\ref{eq:chi_decomp}).
This connection elevates the Bures metric from a formal information-geometric
quantity to a physically measurable response function in mesoscopic Dirac
systems, and provides a concrete prediction for Aharonov--Bohm
interferometry experiments.

These results place the analysis on a clear statistical-mechanical footing:
the flux-tuned Dirac system provides an analytically tractable model in
which the interplay between gauge fields, quantum geometry, and
susceptibility divergences can be worked out exactly within the projected
theory, without relying on semiclassical approximations or perturbation
theory. The exact result $\chi(m) = \pi/(8m)$ constitutes a rare
closed-form expression for an integrated geometric susceptibility in a
relativistic quantum system, and may serve as a benchmark for more complex
interacting or disordered Dirac systems where exact results are
unavailable. The framework developed here can be extended to
multi-parameter deformations, interacting systems, and curved backgrounds,
where richer geometric structures may arise.

For realistic Dirac materials such as graphene on hBN
substrates~\cite{Giovannetti2007} or Kekul\'e-distorted
lattices~\cite{Gomes2012}, induced gaps typically lie in the meV range,
corresponding to mass parameters $m$ for which the predicted Lorentzian
peak $g_{\lambda\lambda} = m^2/[4(m^2 + \nu^2)^2]$ is narrow and strongly
enhanced. In mesoscopic graphene rings, flux variations corresponding to a
single flux quantum are experimentally accessible~\cite{Russo2008}, placing
the relevant range of $\nu$ well within the regime where the Lorentzian
profile dominates. For $m \sim 1\,\mathrm{meV}$ and $R \sim 1\,\mu\mathrm{m}$,
the peak geometric susceptibility $g_{\lambda\lambda}^{\max} = 1/(4m^2)$
corresponds to a persistent current susceptibility of order $(e/h)^2/m$,
which is in principle detectable in existing Aharonov--Bohm interferometry
setups. The integrated susceptibility $\chi(m) = \pi/(8m)$ provides a
direct quantitative prediction for the total geometric response accumulated
over a single flux period, which could be extracted from measurements of
persistent current susceptibility or orbital magnetization as a function of
applied flux. These considerations suggest that the predicted geometric
response may be observable in existing experimental setups, providing a
possible route toward probing information-geometric susceptibilities in
Dirac systems.

\section*{Acknowledgments}

The author would like to express their sincere gratitude to the Conselho
Nacional de Desenvolvimento Cient\'ifico e Tecnol\'ogico (CNPq), and
Funda\c{c}\~ao Cearense de Apoio ao Desenvolvimento Cient\'ifico e
Tecnol\'ogico (FUNCAP) for their valuable support. He is supported by grants
No. 309553/2021-0 (CNPq), 420854/2025-8 (CNPq) and by Project
UNI-00210-00230.01.00/23 (FUNCAP).

\section*{Declaration of Generative AI in Scientific Writing}

The author used a generative AI tool solely for language refinement and clarity
improvement. All scientific content, derivations, analysis, and conclusions are
entirely the responsibility of the author.

\section*{Conflicts of Interest}

The author declares that there is no conflict of interest in this manuscript.

\section*{Data Availability Statement}

No data was used for the research described in this article.


\end{document}